\documentclass[11pt,a4paper]{article}

\usepackage[T1]{fontenc}
\usepackage[utf8]{inputenc}
\usepackage{lmodern}
\usepackage[numbers]{natbib}
\usepackage{amsmath}
\usepackage{amssymb}
\usepackage{array}
\usepackage{booktabs}
\usepackage{enumitem}
\usepackage{float}
\usepackage{graphicx}
\usepackage{tikz}
\usepackage{xurl}
\usepackage[margin=1in]{geometry}
\IfFileExists{microtype.sty}{\usepackage{microtype}}{}
\usepackage[hidelinks]{hyperref}
\usetikzlibrary{arrows.meta,positioning,shapes.geometric}

\makeatletter
\g@addto@macro\UrlBreaks{\do\_\do\-\do\.\do\/\do\=}
\makeatother
\DeclareRobustCommand{\param}[1]{\texttt{\nolinkurl{#1}}}

\title{Evaluating Local Language Model Agents for Reproducible Data Engineering:\\ An Empirical Software Engineering Study of Mobility Workflows}

\author{%
\parbox{0.94\textwidth}{\centering
Jorge García-Carrasco\textsuperscript{1,*}\quad
Javier Sanchis\textsuperscript{1}\quad
Alejandro Reina-Reina\textsuperscript{1}\quad
Alejandro Maté\textsuperscript{1}\quad
Juan Trujillo\textsuperscript{1}\\[1.2ex]
\textsuperscript{1}Lucentia Research, Department of Software and Computing Systems, University of Alicante,\\
Ctra. de Sant Vicent del Raspeig s/n, 03690 Sant Vicent del Raspeig, Alicante, Spain\\[0.8ex]
\textsuperscript{*}Corresponding author: \texttt{jorge.g@ua.es}\\[0.4ex]
\texttt{javier.sanchis@ua.es}, \texttt{alejandro.reina@ua.es},
\texttt{amate@dlsi.ua.es}, \texttt{jtrujillog@dlsi.ua.es}\\[1.2ex]
{\small Preprint. Submitted to \emph{Information and Software Technology} (Elsevier), manuscript
number INFSOF-D-26-01448; currently under review. Replication package:
\url{https://doi.org/10.5281/zenodo.21397610}.}
}}

\date{}

\begin{document}

\maketitle

\begin{abstract}
\textbf{Context:} Large language model (LLM) agents are increasingly used as software and data-engineering assistants, yet evidence about locally deployable open-weight agents remains limited. Existing evaluations often emphasize textual responses, isolated code generation, or generic tool calls rather than the validity of complete engineering artifacts.

\textbf{Objectives:} We evaluate whether local LLM agents can produce correct and reproducible data-engineering artifacts, quantify the effect of a closed-loop workspace condition, and examine trade-offs involving model scale, architecture, quantization, runtime, tool use, and failure behavior.

\textbf{Methods:} We introduce a benchmark of fifteen tasks using mobility workflows as a domain-grounded testbed. Tasks cover data discovery, connector construction, transport-feed processing, semantic enrichment, feature engineering, validation, visualization, and reporting. Deterministic checkers assess generated scripts, tables, structured files, figures, and reports. Ten local model configurations are evaluated in one-shot and closed-loop workspace conditions, with five repetitions per model, mode, and task, yielding 1,500 scored attempts on a consumer-grade GPU.

\textbf{Results:} Among models larger than two billion parameters, the workspace condition increases pass rates by 26.7--52.0 percentage points over one-shot generation. The strongest configuration reaches 85.3\% artifact-level success. A quantized 9-billion-parameter model reaches 69.3\% with an approximately 6.5~GB memory footprint. Gains are largest when intermediate artifacts expose errors that the agent can inspect and repair.

\textbf{Conclusion:} Local open-weight agents can support a meaningful subset of software-intensive data-engineering work, but their reliability depends on model capability, task verifiability, and deterministic validation. The benchmark provides a reproducible method for evaluating complete agent configurations before they are adopted in engineering workflows.
\end{abstract}


\medskip
\noindent\textbf{Highlights.}
\begin{itemize}\setlength{\itemsep}{2pt}\setlength{\parskip}{0pt}
\item Artifact checks evaluate local LLM agents on 15 data-engineering tasks.
\item Closed-loop workspaces improve pass rates by up to 52 points.
\item The strongest local configuration reaches 85.3
\item A quantized 9B model reaches 69.3
\item Execution traces expose reliability, latency, and failure trade-offs.
\end{itemize}

\section{Introduction}
\label{sec:introduction}

Large language models (LLMs) have become software-engineering assistants for activities such as code generation, code review, data preparation, validation, and documentation \citep{hollmann2023large,jiang2026survey,feldt2023towards,pornprasit2024codereview,nguyen2025codesemantics}. Their engineering value, however, depends on more than producing plausible text or isolated snippets. A generated script must execute, a structured file must satisfy its schema, and a derived result must remain consistent with its inputs. These requirements shift evaluation from response plausibility toward the validity of the engineering artifacts produced by the model.

When an LLM is embedded in a system that can plan, inspect a workspace, invoke tools, observe execution results, and revise its outputs, it is commonly described as an LLM-based agent or tool-using language model \citep{yao2023react,schick2023toolformer,wang2024survey}. Recent software-engineering research has begun to examine multi-agent generation of structured requirements and the validity of artifacts produced under different agent configurations \citep{liu2026ucdllm,kim2026artifactvalidity}. Nevertheless, evidence remains limited for complete, locally deployable agent workflows evaluated through heterogeneous artifacts rather than a single code-generation task.

Hosted frontier models are useful capability baselines, but many public-sector, research, and engineering organizations also need deployment options that keep data within organizational boundaries and provide control over model versions, inference settings, availability, and cost. Local open-weight models offer these properties, but their smaller size and quantization may affect their ability to operate a workspace reliably. An empirical assessment must therefore consider not only final correctness, but also model scale, resource use, interaction behavior, and failure modes.

Data engineering provides a software-intensive setting for this assessment. Constructing a data workflow involves programming, schema conformance, testing, provenance, domain rules, validation, and reproducible reporting. Mobility workflows are a particularly useful testbed because they combine statistical indicators, catalog records, public-transport feeds, geospatial points of interest, executable transformations, and derived analytical artifacts. Correctness requires both general engineering behavior and domain-specific care, including preserving provenance, filtering decoy records, handling GTFS service-day times beyond \texttt{24:00:00}, computing geospatial distances in meters, and avoiding unsupported claims in reports.

This paper presents a reproducible benchmark and controlled empirical study of local LLM agents as data-engineering assistants. The fifteen tasks follow five stages of a mobility workflow: discovering and connecting to data, processing public-transport feeds, enriching records with semantic data, engineering features and analytics, and validating and reporting results. Each task specifies input fixtures and expected output artifacts. Deterministic checkers then parse, execute, compare, or inspect those artifacts against schemas, domain rules, numerical constraints, and report requirements. The harness records success, failure category, runtime, and tool interaction, enabling the study to examine both artifact validity and the engineering behavior that precedes submission.

The work makes three contributions:
\begin{enumerate}
    \item A domain-grounded artifact benchmark with fixed fixtures, deterministic checkers, machine-readable traces, and a public replication package.
    \item A controlled study of one-shot and closed-loop conditions across ten local model configurations, fifteen tasks, and five repetitions per condition.
    \item Empirical evidence about artifact correctness, task verifiability, model scale, architecture, quantization, latency, tool interaction, and failure behavior, with implications for validating LLM-assisted engineering workflows.
\end{enumerate}

The evaluation is organized around four research questions:
\begin{enumerate}
    \item[(RQ1)] How does artifact success differ between one-shot generation and a closed-loop workspace condition in which a local LLM can inspect inputs, execute code, and revise its artifacts?
    \item[(RQ2)] What level of artifact correctness and reliability do locally deployable open-weight models achieve on the mobility data-engineering tasks?
    \item[(RQ3)] How do artifact success, runtime, and tool interactions vary with model size and architecture, and what engineering trade-offs emerge?
    \item[(RQ4)] How do task success and computational efficiency differ between aggressive post-training quantization and unquantized BF16 for the model sizes available in both formats?
\end{enumerate}

The experimental scope is deliberately restricted to models that can be served on a single 24~GB consumer GPU. This keeps the study aligned with locally controlled engineering environments and permits a controlled comparison of model scale, architecture, and precision within one serving stack. The benchmark is also intended as a reusable evaluation instrument: future agent configurations can be assessed under the same fixtures and checkers, while organizations can substitute representative local tasks before adopting an LLM-based engineering assistant. The benchmark does not claim that mobility tasks cover all software or data-engineering practice; rather, they provide a heterogeneous and reproducible testbed for studying artifact-producing agent workflows.

The remainder of the paper is organized as follows. Section~\ref{sec:related-work} reviews LLM agents, LLM-assisted software engineering, executable benchmarks, local inference, reproducibility, and mobility data. Section~\ref{sec:benchmark-design} describes the fixtures, task suite, interaction modes, and scoring rules. Section~\ref{sec:experimental-setup} presents the model and execution configuration. Section~\ref{sec:results} reports the empirical results, Section~\ref{sec:discussion} discusses their engineering implications and threats to validity, and Section~\ref{sec:conclusion} concludes the paper.

\section{Related work}
\label{sec:related-work}

This paper studies locally deployable LLM agents as software and data-engineering assistants, with success measured from reproducible artifacts rather than textual plausibility. We therefore connect research on LLM-assisted software engineering with work on agents, executable benchmarks, tool-use evaluation, efficient local inference, reproducibility, and open mobility data.

A first strand of related work studies how language models can act through external tools or environments. ReAct introduced a prompting paradigm in which language models interleave reasoning traces with task-directed actions, allowing models to update plans and gather external information during problem solving \citep{yao2023react}. Toolformer showed that language models can be trained to decide when and how to call external APIs, including calculators, retrieval systems, translation systems, and calendars \citep{schick2023toolformer}. These works motivate the comparison between one-shot generation and tool-mediated interaction to determine whether tool-usage improves performance in tasks related to mobility data engineering.

Broader agent benchmarks evaluate language models in interactive environments. AgentBench proposed a multi-environment benchmark for assessing language models as agents in open-ended, multi-turn settings \citep{liu2024agentbench}. Such work is important for evaluating planning and decision making, but it typically emphasizes general-purpose agency across heterogeneous environments. In contrast, the present benchmark narrows the scope to reproducible engineering tasks and the validity of submitted artifacts.

Within software engineering, empirical studies have assessed LLMs for code review automation and code-semantic understanding \citep{pornprasit2024codereview,nguyen2025codesemantics}. Other work has moved from single-model assistance toward structured and agentic workflows. UCD-LLM uses a multi-agent framework to generate use-case-diagram requirements \citep{liu2026ucdllm}, while Kim compares artifact validity under varying LLM-agent configurations \citep{kim2026artifactvalidity}. These studies establish direct precedent for evaluating LLM configurations through software-engineering outcomes. The present study extends this perspective to complete data-engineering artifacts, a local serving constraint, heterogeneous task families, repeated execution, and failure analysis.

A separate body of work has established executable checking as the central evaluation principle for code generation. The Codex paper introduced HumanEval, a benchmark of Python programming problems evaluated by functional correctness rather than surface similarity \citep{chen2021codex}. MBPP extended evaluation to short natural-language programming tasks intended to be solvable by entry-level programmers \citep{austin2021programsynthesis}. CodeXGLUE broadened evaluation across code understanding and generation tasks, providing a multi-task benchmark for program-related machine learning \citep{lu2021codexglue}. More recent benchmarks move closer to realistic engineering work: DS-1000 evaluates data-science code generation using problems derived from practical Python library usage \citep{lai2023ds1000}, and SWE-bench evaluates whether language models can resolve real GitHub issues by editing existing repositories and passing associated tests \citep{jimenez2024swebench}. These benchmarks are especially relevant because they use executable or repository-level evaluation. However, they primarily target programming solutions or repository patches. The present benchmark instead evaluates a heterogeneous set of connected engineering outputs, including executable transformations, cleaned tables, validation reports, semantic-linking files, analytical matrices, figures, and reproducible reports. This broadens artifact-level evaluation from source code alone to the products of a complete data-engineering workflow.

Another line of research focuses directly on tool-using models. API-Bank introduced a benchmark and runnable evaluation system for tool-augmented language models, covering API planning, retrieval, and invocation \citep{li2023apibank}. StableToolBench addressed the instability of live online APIs by introducing a virtual API server and a more stable evaluation process for large-scale tool-learning benchmarks \citep{guo2024stabletoolbench}.  Function-calling benchmarks further refine how tool calls are checked. The Berkeley Function Calling Leaderboard evaluates serial, parallel, and multi-step function calling using structured checking, including abstract-syntax-tree-based evaluation \citep{patil2025bfcl}. $\tau$-bench evaluates tool-agent-user interaction in simulated real-world domains and introduces pass$^k$ as a measure of behavioral reliability across repeated trials \citep{yao2024taubench}. ToolSandbox evaluates stateful conversational tool use with intermediate and final milestones, emphasizing state dependencies, canonicalization, and insufficient-information cases \citep{lu2025toolsandbox}. In this context, we equip LLMs with a minimal tool suite to execute data-engineering workflows and compare the complete closed-loop workspace condition with a one-shot baseline.

The present work is also related to research on efficient local language-model inference. In this sense, low-bit post-training quantization methods such as GPTQ \citep{frantar2023gptq} and AWQ \citep{lin2024awq} reduce memory and inference costs, making larger models more practical on local hardware. This paper does not propose a new quantization or inference method; instead, it treats local serving, model size, architecture, and precision as configuration choices in an empirical evaluation of artifact-producing engineering assistants.

Reproducibility is also central to empirical artificial intelligence and software engineering. Many Artificial Intelligence (AI) papers did not document enough information about experiments, data, and methods to support reproducibility \citep{gundersen2018reproducibility}. Reproducibility practices introduced around the NeurIPS reproducibility program, including checklists, code submission, and reproducibility challenges, were systematized shortly thereafter \citep{pineau2021reproducibility}. DataPerf extended benchmark thinking toward data-centric AI, emphasizing that datasets and data-development processes themselves require systematic evaluation \citep{mazumder2023dataperf}. The benchmark in this paper adopts these concerns operationally: inputs are fixed, tasks are versioned, outputs are checked deterministically, and logs are machine-readable.

Finally, the domain layer of this benchmark builds on public mobility and open-data infrastructures, which provide both standardized data formats and realistic sources for data-oriented tasks. On the mobility side, GTFS defines a widely used tabular standard for public-transport schedules, including files such as \texttt{stops.txt}, \texttt{trips.txt}, and \texttt{stop\_times.txt}. Its specification also defines service-day time semantics, under which valid stop times may exceed \texttt{24:00:00} \citep{gtfsScheduleReference2026}. Together with the MobilityData GTFS validator, which offers a reference point for transit-feed quality rules and validation terminology \citep{mobilityDataValidator2026}, GTFS provides a suitable basis for benchmark tasks involving parsing, validation, cleaning, and service-intensity computation.

Beyond mobility feeds, the benchmark incorporates public statistical, catalog, and semantic-data sources in order to cover broader urban-data workflows. Eurostat supports programmatic access to statistical data and metadata through public APIs, making it relevant for city-level indicator tasks \citep{eurostatApi2026}. Similarly, the data.europa.eu portal documents API access to catalog metadata, enabling dataset discovery and metadata extraction tasks \citep{dataEuropaApi2026}. Finally, Wikidata provides a SPARQL query service for structured semantic retrieval, which supports point-of-interest enrichment and geospatial linking tasks \citep{wikidataSparql2026}.

Taken together, the related work surveyed above paints a clear trajectory: language-model agents are becoming increasingly capable at tool use, code generation, and interactive problem solving, while local deployment and reproducible evaluation methods are maturing in parallel. These advances open a broad space of domain-specific applications that remain largely unexplored. This paper enters that space by asking how complete local agent configurations perform when they must produce validated engineering artifacts rather than plausible responses. Mobility supplies heterogeneous data sources and domain conventions, but the empirical concern is broader: how workspace feedback, model capability, resource constraints, and task verifiability affect LLM-assisted engineering. To the best of our knowledge, no prior work combines artifact-level mobility data-engineering tasks with a controlled local-model sweep that measures workspace-condition gain, runtime, tool interaction, quantization sensitivity, and failure categories.

\section{Benchmark design}
\label{sec:benchmark-design}

The benchmark operationalizes an LLM-based data-engineering assistant as a system that transforms a natural-language task specification and bounded workspace into verifiable output artifacts. Each task provides an instruction, relevant input fixtures, and a specification of the files to be produced. Success is determined from the artifacts rather than from the model's explanation: deterministic checkers parse, execute, compare, or inspect outputs and verify schemas, numerical constraints, domain rules, and report requirements.

This design reflects software-engineering practice for data-intensive systems. A useful assistant must do more than describe a plausible workflow: it must select valid records, understand the actual data layout, write executable transformations or structured outputs, handle domain conventions, and leave artifacts that can be tested and audited. The tasks follow an urban analytics workflow from data discovery and feed processing through enrichment, feature construction, validation, and reporting. Decomposing that workflow into independently checked tasks permits capability-level diagnosis while preserving the relationship between upstream transformations and downstream analytical artifacts.

The data used by the benchmark is drawn from public mobility and urban-data sources. The Eurostat-style component provides city-level indicator metadata and values for mobility, population, environment, and living-condition variables \citep{eurostatApi2026}. The catalog component provides data.europa.eu-style records, including relevant mobility datasets and non-mobility decoys that test whether a model can distinguish source records from distractors \citep{dataEuropaApi2026}. The transport component contains a sampled GTFS feed from a public mobility catalog, which supports tasks involving stops, trips, stop times, service-day semantics, and feed validation \citep{mobilityDataCatalog2026}. The semantic component contains Wikidata-style point-of-interest records for cultural-accessibility enrichment and geospatial linking \citep{wikidataSparql2026}.

For reproducibility, these sources are not queried live during evaluation. Instead, all benchmark inputs are committed as local fixture snapshots and every model receives the same files, prompts, workspace structure, and scoring logic. This avoids external variability from API outages, schema changes, or updated source records, making it possible to attribute differences in performance to the model and interaction mode rather than to changes in the data environment.

\subsection{Task suite}
\label{sec:task-suite}

The benchmark comprises fifteen tasks that correspond to recurring stages in mobility data-engineering workflows: discovering relevant datasets, ingesting and cleaning transport feeds, enriching records with semantic data, computing analytical features, and validating/reporting results. Table~\ref{tab:tasks} summarizes the five task families.

\begin{table}[htbp]
\centering
\caption{Benchmark task families.}
\label{tab:tasks}
\begin{tabular}{lll}
\toprule
Family & Task IDs & Scope \\
\midrule
A & T01--T03 & Dataset discovery and connector setup \\
B & T04--T06 & GTFS processing \\
C & T07--T09 & Semantic enrichment with Wikidata-style data \\
D & T10--T13 & Feature engineering and analytics \\
E & T14--T15 & Validation, visualization, and reporting \\
\bottomrule
\end{tabular}
\end{table}

A description of the fifteen tasks is presented below, grouped by family. For each task, the description states the objective, the requirement specifies what the agent must produce, and the correctness criterion describes how the deterministic checker validates the output.

\textbf{Family A: Dataset discovery and connector setup (T01--T03).} These tasks test whether a model can identify relevant urban indicators, build a data loader, and reason about catalog records to extract mobility entries while excluding decoys, i.e., catalog records that look structurally valid but describe non-mobility datasets and should therefore be filtered out.
\begin{enumerate}
    \item[\textbf{T01.}] \textbf{Indicator selection.} \textit{Requirement:} Select Eurostat urban indicators relevant to mobility and livability and produce a CSV that lists each selected indicator with its theme, a brief justification, and the expected analytical direction. \textit{Correctness:} The output must contain the required columns (theme, justification, direction) and every selected indicator must belong to an approved mobility-related theme.
    \item[\textbf{T02.}] \textbf{Loader construction.} \textit{Requirement:} Build an executable Python loader for the selected Eurostat-style indicator values and produce a normalized long-format CSV. \textit{Correctness:} The script must run without errors and the output CSV must contain the expected indicator, city, year, and value columns with properly melted structure.
    \item[\textbf{T03.}] \textbf{Catalog filtering.} \textit{Requirement:} Extract mobility-related records from catalog metadata and produce a mobility catalog CSV that excludes non-mobility decoy records. \textit{Correctness:} The output must contain all mobility entries and no decoys.
\end{enumerate}

\textbf{Family B: GTFS processing (T04--T06).} These tasks test parsing, validation, data cleaning, and transit-service aggregation, including the domain-specific distinction between invalid clock times and valid extended GTFS service-day times.
\begin{enumerate}[resume]
    \item[\textbf{T04.}] \textbf{Feed validation.} \textit{Requirement:} Validate and parse a GTFS feed and produce a JSON summary of mandatory files, required columns, and row counts. \textit{Correctness:} The JSON must list every mandatory file, every required column per file, and accurate row counts.
    \item[\textbf{T05.}] \textbf{Data cleaning.} \textit{Requirement:} Clean GTFS coordinates, duplicate stops, missing names, and service-day times, producing a cleaning report together with cleaned stops and stop-times tables. \textit{Correctness:} The cleaned tables must remove duplicates and blanks, correct invalid coordinates, and preserve the valid extended service-day time (\texttt{25:10:00}).
    \item[\textbf{T06.}] \textbf{Service intensity.} \textit{Requirement:} Compute stop-level service intensity and produce a stop table with departure counts, route counts, and first/last departure times. \textit{Correctness:} Aggregated values must match the ground-truth schedule and times must respect GTFS service-day semantics.
\end{enumerate}

\textbf{Family C: Semantic enrichment (T07--T09).} These tasks test whether a model can express a plausible Wikidata SPARQL query, clean geospatial POI records, and perform a nearest-stop join using distances in meters.
\begin{enumerate}[resume]
    \item[\textbf{T07.}] \textbf{SPARQL query generation.} \textit{Requirement:} Generate a Wikidata SPARQL query for cultural points of interest. \textit{Correctness:} The query must request entities, labels, coordinates, and location constraints with syntactically valid SPARQL structure.
    \item[\textbf{T08.}] \textbf{POI cleaning.} \textit{Requirement:} Clean Wikidata-style POI records and produce a table with unique entities, labels, types, coordinates, and source. \textit{Correctness:} Duplicate entity identifiers must be deduplicated, blank labels must be handled, and the schema must match the specification.
    \item[\textbf{T09.}] \textbf{Nearest-stop linking.} \textit{Requirement:} Link each POI to its nearest public-transport stop and produce a POI-to-stop table with haversine-style distances in meters. \textit{Correctness:} Distances must be computed in meters (not coordinate degrees) and each POI must be matched to the correct nearest stop.
\end{enumerate}

\textbf{Family D: Feature engineering and analytics (T10--T13).} These tasks test formula-following, pivoting, missingness analysis, and a fixed clustering pipeline.
\begin{enumerate}[resume]
    \item[\textbf{T10.}] \textbf{Accessibility scoring.} \textit{Requirement:} Compute a cultural accessibility score combining distance, departures, routes, and a normalized score. \textit{Correctness:} The formula must be implemented exactly as specified and the output table must contain all required columns.
    \item[\textbf{T11.}] \textbf{Analytical matrix.} \textit{Requirement:} Create a city-year by indicator matrix from long-format Eurostat values. \textit{Correctness:} The matrix must have the correct shape, indices, and populated cells.
    \item[\textbf{T12.}] \textbf{Missingness filtering.} \textit{Requirement:} Select robust indicators based on missingness and produce a missingness report together with a selected matrix using a 70\% coverage rule. \textit{Correctness:} Only indicators with at least 70\% non-missing coverage may be retained.
    \item[\textbf{T13.}] \textbf{City clustering.} \textit{Requirement:} Cluster cities into mobility and livability profiles and produce cluster assignments together with cluster-profile summaries. \textit{Correctness:} Cluster labels must match the ground-truth assignments obtained from the specified algorithm and parameters.
\end{enumerate}

\textbf{Family E: Validation, visualization, and reporting (T14--T15).} These tasks test whether the model can validate the resulting pipeline artifacts and write a compact report that cites generated outputs without unsupported claims.
\begin{enumerate}[resume]
    \item[\textbf{T14.}] \textbf{Output validation.} \textit{Requirement:} Generate an executable validation script for pipeline outputs and produce a validation-summary JSON. \textit{Correctness:} The script must run without errors and the JSON must accurately reflect which checks passed and which failed.
    \item[\textbf{T15.}] \textbf{Reproducible reporting.} \textit{Requirement:} Produce a reproducible Markdown report together with two non-empty analysis figure files. \textit{Correctness:} The report must reference generated artifacts and avoid unsupported claims, while the two required figure files must exist and be non-empty.
\end{enumerate}

Together, these five families cover the full data-engineering lifecycle from discovery and ingestion through cleaning, enrichment, analytics, validation, and reporting. By decomposing the workflow into independently scored tasks, the benchmark pinpoints exactly which stages are within reach of a given model and which remain difficult.

\subsection{Interaction modes}
\label{sec:interaction-modes}

The harness evaluates each task in two modes because one of the main research questions is not only whether a model can solve a task, but whether equipping it with a minimal local workspace loop materially improves its ability to produce correct and reproducible artifacts. In other words, the benchmark asks whether the agentic closed-loop behavior described above, in which a model inspects files, executes code, and repairs its own outputs, translates into measurable engineering value. Comparing a tool-free one-shot condition with a transparent tool-mediated condition allows us to estimate the practical contribution of this deployable agent setup.

We therefore evaluate two modes. In \texttt{chat-only} mode, the model receives the task prompt and output requirements and must provide the required artifacts in a single response. The harness extracts files from the response, writes them into the workspace, and invokes the checker. This mode measures the model's ability to synthesize complete outputs without local inspection or execution.

In \texttt{agent-with-tools} mode, the model operates in an isolated local workspace. The tool interface exposes five actions: reading a file, writing a file, editing a string in an existing file, running a shell command, and submitting the final answer. This minimal set of tools allows the model to inspect inputs, create scripts, run local checks, view command output, revise files, and decide when to submit. The hidden checker is invoked only after submission. The interface is deliberately minimal so that the benchmark measures the interaction between model capability and a transparent workspace loop, rather than the behavior of a complex external agent framework.

Both modes receive the same task definition and are scored with the same checker. The difference in pass rate between \texttt{agent-with-tools} and \texttt{chat-only} is reported as the ``agentic gain''. This quantity measures the complete workspace condition---including tool access, iterative interaction, execution feedback, artifact inspection, and cumulative context---rather than the isolated effect of tools.

\subsection{Evaluation and failure categories}
\label{sec:evaluation}

Each task has a deterministic checker that returns a binary success value, checker errors, a failure category, and task-specific metadata. The primary evaluation focuses on artifact correctness, because this is the property that matters most in data-engineering automation: generated files must have the required structure, scripts must execute, etc. The checkers therefore evaluate the artifact type appropriate to the task, including schemas, row-level rules, JSON fields, etc. For artifacts containing natural language, the checker evaluates structural compliance, required coverage, and consistency with generated data, but does not attempt to score subjective prose quality. All generated artifacts and logs are retained, so future analyses can apply human review or LLM-judge methods to prose quality if that question is of interest, but the benchmark's primary score deliberately remains tied to reproducible artifact-level correctness.

We group unsuccessful attempts into the following categories:
\begin{enumerate}
    \item \textbf{Logic errors:} Structurally valid outputs that violate task rules, such as using the wrong filter, computing the wrong aggregate, or returning an incorrect row count.
    \item \textbf{Protocol errors:} Invalid or missing tool calls, including malformed responses; an attempt is stopped after eight consecutive protocol errors.
    \item \textbf{Format errors:} Files that exist but cannot be parsed, such as malformed CSV or invalid JSON.
    \item \textbf{Schema errors:} Missing files, missing columns, or unusable artifact structure, for example omitting a required result CSV or report section.
    \item \textbf{Wrong data source:} Outputs that use identifiers or records not present in the provided fixture, such as selecting a decoy file from the catalog.
    \item \textbf{Tool-error loops:} Repeated failed tool calls or repeated commands that do not make progress; the harness stops an attempt after eight repeated failures of the same tool call.
    \item \textbf{Tool loops:} Repeated valid tool calls that do not make progress; the harness stops an attempt after twenty repetitions of the same tool call.
    \item \textbf{Domain errors:} Violations of mobility-specific conventions, such as incorrect GTFS time handling, geospatial matching, or query constraints.
\end{enumerate}
These categories allow the analysis to distinguish a model that cannot produce the required file shape from a model that produces valid files but applies the wrong domain logic.

Every reported metric is extracted automatically from the execution trace rather than assigned through subjective human judgment. The run logger records model identifier, task identifier, interaction mode, repetition, final checker result, failure category, runtime, and tool-call counts. This structured logging makes the evaluation fully reproducible and enables direct comparison across models, tasks, and families.

\section{Experimental setup}
\label{sec:experimental-setup}

The experiments evaluate Qwen-family local models \citep{qwen3.6-27b} served through a local \texttt{llama.cpp} \citep{llamacpp2026} endpoint on a single NVIDIA GeForce RTX 4090 GPU. The evaluated suite spans the Qwen 3.5 and 3.6 families, with model sizes from 2B to 35B parameters in both dense and mixture-of-experts (MoE) architectures \citep{shazeer2017outrageously} (Table~\ref{tab:models}). The 35B-A3B configurations are MoE models with 35B total parameters and approximately 3B active parameters during each forward pass. Because the 27B dense and 35B-A3B MoE models do not fit into the 24~GB of VRAM available on the RTX 4090 in BF16, we decided to use the models quantized with Unsloth Dynamic 2.0 GGUF at UD-Q4\_K\_XL \citep{unsloth_quant}. This is a higher-fidelity 4-bit quantization that uses larger intermediate tensors and dynamic scaling to preserve more accuracy in critical weight groups, while still remaining within the memory budget of consumer GPUs. Using the same quantization across evaluated sizes reduces precision level as a source of variation in the scale comparison. However, to test whether this aggressive quantization harms smaller models that have less representational headroom, additional unquantized BF16 variants of the 2B, 4B, and 9B models were also evaluated because they fit on a consumer-grade GPU.

The model set is deliberately a controlled within-family sweep rather than a broad leaderboard across unrelated open-weight families. This choice prioritizes the questions that are most central to local deployment in this paper: how capability changes with scale, whether a sparse MoE model changes the latency--accuracy trade-off, whether a newer generation improves the frontier, and whether aggressive quantization changes success near the capability threshold. Adding many model families would introduce additional confounds, including tokenizer differences, chat templates, training data, instruction tuning, tool-call behavior, and quantization availability. This focused scope supports a more controlled analysis of scale and efficiency trade-offs, while the released benchmark can be extended to other model families in future replications.

\begin{table}[htbp]
\centering
\caption{List of models evaluated in the benchmark.}
\label{tab:models}
\begin{tabular}{lllll}
\toprule
Model ID & Family & Size & Architecture & Quantization \\
\midrule
qwen35\_2b\_q4 & Qwen 3.5 & 2B & Dense & UD-Q4\_K\_XL \\
qwen35\_2b\_bf16 & Qwen 3.5 & 2B & Dense & BF16 \\
qwen35\_4b\_q4 & Qwen 3.5 & 4B & Dense & UD-Q4\_K\_XL \\
qwen35\_4b\_bf16 & Qwen 3.5 & 4B & Dense & BF16 \\
qwen35\_9b\_q4 & Qwen 3.5 & 9B & Dense & UD-Q4\_K\_XL \\
qwen35\_9b\_bf16 & Qwen 3.5 & 9B & Dense & BF16 \\
qwen35\_27b\_q4 & Qwen 3.5 & 27B & Dense & UD-Q4\_K\_XL \\
qwen35\_35b\_a3b\_q4 & Qwen 3.5 & 35B & MoE & UD-Q4\_K\_XL \\
qwen36\_27b\_q4 & Qwen 3.6 & 27B & Dense & UD-Q4\_K\_XL \\
qwen36\_35b\_a3b\_q4 & Qwen 3.6 & 35B & MoE & UD-Q4\_K\_XL \\
\bottomrule
\end{tabular}
\end{table}

All reported results use the Unsloth-recommended generation configuration for Qwen 3.5 non-thinking precise-coding mode: \param{temperature=0.6}, \param{top_p=0.95}, \param{top_k=20}, \param{min_p=0.0}, \param{presence_penalty=0.0}, and \param{repeat_penalty=1.0}. The choice was also motivated by preliminary observations with greedy decoding (\param{temperature=0.0}), where repetitions were nearly identical and some models entered repetitive tool loops that did not converge. These preliminary runs were used to select a stable common configuration and are not included in the 1,500 scored attempts. The study therefore evaluates the reported models under one fixed stochastic generation policy; it does not attempt to estimate the independent effect of decoding parameters.

Building on the offline-fixture design described above, all benchmark runs use the same committed snapshot so that every model is evaluated under identical input conditions. For the evaluation reported here, this snapshot contains 5,373 Eurostat value rows, 160 Eurostat indicator descriptions, 26 catalog records including one selected GTFS feed record, a GTFS subset with 81 stops after controls, 80 trips, 180 stop-time rows, and 51 Wikidata POI rows after controls. This fixed snapshot operationalizes the reproducibility requirement: the benchmark does not depend on live APIs that might fail, return inconsistent results, or change schema between runs. As a result, input data are not a source of variation when comparing model performance. The raw source downloads are retained outside the public benchmark input tree, while the benchmark inputs and provenance record are versioned with the harness.

Known defects are also manually inserted into the local fixtures to make validation and cleaning measurable. The controlled defects include duplicate GTFS stop identifiers, a blank GTFS stop name, invalid GTFS coordinates, a valid extended GTFS service-day time of \texttt{25:10:00}, a duplicate Wikidata entity identifier, and a blank Wikidata label. These cases are documented in the benchmark provenance metadata and are used only to evaluate whether models handle realistic data-cleaning conditions. Including defects is essential because a benchmark in which all data are pristine would reward models that simply copy inputs without exercising error-handling logic.

Finally, agent-mode execution is implemented as a deliberately small closed-loop harness rather than as a full external agent framework. The purpose of this design is to expose only the workspace operations needed for data engineering while keeping the source of any performance gain interpretable. Each attempt proceeds as follows:
\begin{enumerate}
    \item The harness creates an isolated attempt workspace containing the fixed input snapshot and empty output, source, report, and figure directories.
    \item The model receives the task prompt together with a system prompt specifying five JSON-schema tools: \param{read}, \param{write}, \param{edit}, \param{bash}, and \param{submit}.
    \item At each turn, the model emits at most one tool call. The harness validates the call, executes it locally, records the event, and returns the result to the model as the next observation.
    \item The model may inspect files, create or edit scripts, execute local checks, and repair its artifacts using only the visible workspace evidence.
    \item When the model calls \param{submit}, the interaction stops and the hidden deterministic checker evaluates the generated artifacts. The checker result is not available to the model during the attempt.
\end{enumerate}
To prevent agents from bypassing the benchmark by inspecting hidden solutions, checker code, or files outside the task environment, each attempt workspace contains only the initial task inputs and the directories needed for generated artifacts. File access and shell commands are constrained to this workspace, network access is disabled, and checker code and reference expectations are not exposed to the model during the attempt.

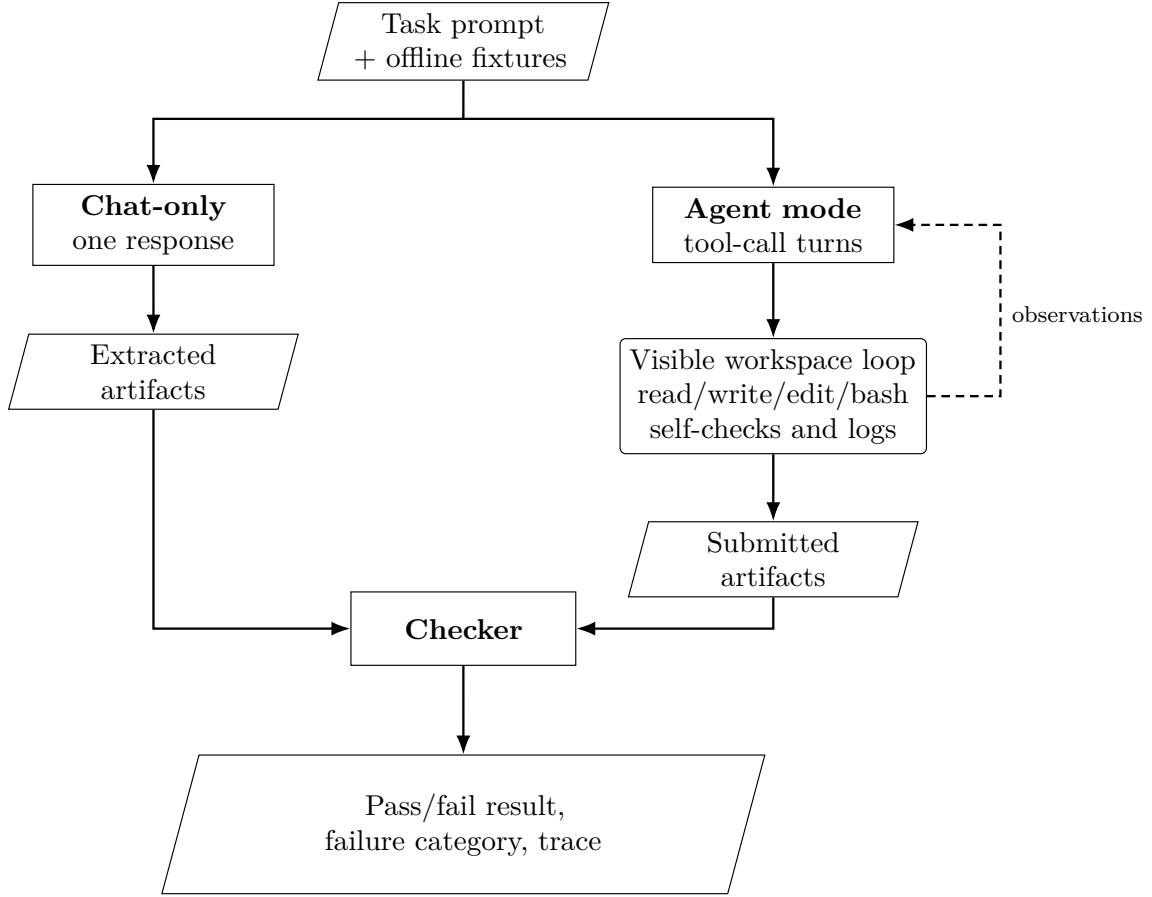
\begin{figure}[H]
\centering
\resizebox{0.96\textwidth}{!}{%
\begin{tikzpicture}[
    font=\small,
    lane/.style={draw=black, text=black, align=center, minimum height=8mm, text width=27mm},
    loop/.style={draw=black, rounded corners=2pt, text=black, align=center, minimum height=10mm, text width=35mm},
    io/.style={draw=black, text=black, trapezium, trapezium left angle=75, trapezium right angle=105, align=center, minimum height=8mm, text width=28mm},
    checker/.style={draw=black, text=black, align=center, minimum height=9mm, text width=25mm},
    score/.style={draw=black, text=black, trapezium, trapezium left angle=75, trapezium right angle=105, align=center, minimum height=17mm, text width=36mm},
    flow/.style={draw=black, -Latex, line width=0.8pt},
    feedback/.style={draw=black, -Latex, densely dashed, line width=0.8pt}
]
\node[io] (input) at (0,0) {Task prompt\\+ offline fixtures};

\node[lane] (chat) at (-3.8,-2.25) {\textbf{Chat-only}\\one response};
\node[lane] (agent) at (3.8,-2.25) {\textbf{Agent mode}\\tool-call turns};

\node[io] (chatout) at (-3.8,-4.05) {Extracted\\artifacts};
\node[loop] (toolloop) at (3.8,-4.35) {Visible workspace loop\\read/write/edit/bash\\self-checks and logs};
\node[io] (agentout) at (3.8,-6.35) {Submitted\\artifacts};

\node[checker] (eval) at (0,-7.20) {\textbf{Checker}};
\node[score] (scores) at (0,-9.60) {Pass/fail result,\\failure category, trace};

\draw[flow] (input.south) -- (0,-0.95) -| (chat.north);
\draw[flow] (0,-0.95) -| (agent.north);
\draw[flow] (chat) -- (chatout);
\draw[flow] (agent) -- (toolloop);
\draw[feedback] (toolloop.east) -- ++(0.9,0) |- node[pos=0.24, right, font=\scriptsize] {observations} (agent.east);
\draw[flow] (toolloop) -- (agentout);
\draw[flow] (chatout.south) |- (eval.west);
\draw[flow] (agentout.south) |- (eval.east);
\draw[flow] (eval) -- (scores);
\end{tikzpicture}
}
\caption{Comparison between chat-only and agent-mode evaluation. Both modes receive the same prompt and offline fixtures, produce artifacts, and are scored by the same hidden deterministic checker; only agent mode can iterate through visible workspace tools before submission.}
\label{fig:agent-workflow}
\end{figure}

Figure \ref{fig:agent-workflow} makes the distinction between the two evaluated modes more explicit: in \param{chat-only} mode, the model must synthesize all required files in one response without observing the workspace or executing code. In \param{agent-with-tools} mode, the same model can iteratively gather evidence and self-verify before submission, as described above.

Each attempt is also bounded by fixed execution caps to prevent non-converging runs from consuming unbounded time while preserving the failure signal. We allow at most 1000 interaction iterations, 1000 total tool calls, one hour of wall-clock time per task, and 60 seconds for each shell command. To stop traces that are clearly no longer making progress, the harness also terminates after eight consecutive malformed tool-call responses, eight repeated failed tool calls, or 20 repeated identical tool calls. When a cap is reached, the stop reason and cap flags are recorded in the trace and the attempt is evaluated as a failed model run unless the trace indicates an infrastructure problem. The caps allowed agents to recover from local errors and pursue alternative solutions while terminating traces that failed to make progress within the one-hour task budget.

\section{Results}
\label{sec:results}

In this section, we present the results obtained during the benchmark process. We evaluate the ten local Qwen configurations listed in Table~\ref{tab:models} on the fifteen defined tasks in both \param{chat_only} and \param{agent_tools} modes. The model set spans 2B, 4B, 9B, 27B, and 35B-A3B parameter regimes. All size classes are evaluated with UD-Q4\_K\_XL quantization, while the 2B, 4B, and 9B Qwen 3.5 models are also evaluated in BF16 because they fit on the available RTX 4090 consumer-grade GPU. Only the 27B dense and 35B-A3B MoE configurations are evaluated for both Qwen 3.5 and Qwen 3.6, as the smaller variants have not been published in the 3.6 version. Each model/mode/task condition is repeated five times, yielding $10 \times 2 \times 15 \times 5 = 1500$ scored attempts. We report Wilson 95\% intervals as descriptive summaries of pooled attempt-level pass rates. Because attempts are organized within fixed task strata and only five repetitions are available per model/task cell, these intervals should not be interpreted as population-level uncertainty over all possible data-engineering tasks.

\subsection{Main results}

First, we present three main comparisons:
\begin{itemize}
    \item \textbf{Quantized agent-mode analysis:} we analyze the natural operating setting for the benchmark, where all models use the same local workspace interface and the same UD-Q4\_K\_XL quantization level.
    \item \textbf{Workspace-condition gain:} we compare \param{agent_tools} against \param{chat_only} to estimate the observed benefit of the complete closed-loop configuration.
    \item \textbf{Precision sensitivity:} we exploratorily compare Q4 and BF16 for the 2B, 4B, and 9B models available in both formats.
\end{itemize}

We first compare pass rate, median runtime, and median tool calls for quantized models in \param{agent_tools} mode, aggregated across tasks and repetitions (Table~\ref{tab:main-results}). Artifact success rises sharply with scale at the lower end of the sweep. The 2B model passes only 2/75 attempts. The 4B model reaches 37.3\%, while the 9B model reaches 69.3\% and is the first size in this sweep to complete a majority of attempts.

\begin{table}[htbp]
\centering
\caption{Main agent-mode results for quantized models (5 repetitions per task, $N=75$ per model). Pass-rate entries report percentage and Wilson 95\% confidence interval.}
\label{tab:main-results}
\scriptsize
\setlength{\tabcolsep}{2pt}
\begin{tabular*}{\linewidth}{@{\extracolsep{\fill}}l l r r r@{}}
\toprule
Model ID & Arch. & Pass rate & Med.\ runtime (s) & Med.\ tools \\
\midrule
\param{qwen35_2b_q4} & Dense & 2.7\% [0.7, 9.2] & 12.3 & 5 \\
\param{qwen35_4b_q4} & Dense & 37.3\% [27.3, 48.6] & 12.3 & 7 \\
\param{qwen35_9b_q4} & Dense & 69.3\% [58.2, 78.6] & 10.5 & 7 \\
\param{qwen35_27b_q4} & Dense & 74.7\% [63.8, 83.1] & 29.9 & 9 \\
\param{qwen35_35b_a3b_q4} & MoE & 77.3\% [66.7, 85.3] & 9.6 & 7 \\
\param{qwen36_27b_q4} & Dense & 85.3\% [75.6, 91.6] & 31.3 & 8 \\
\param{qwen36_35b_a3b_q4} & MoE & 81.3\% [71.1, 88.5] & 14.7 & 8 \\
\bottomrule
\end{tabular*}
\end{table}

Beyond 9B, pass rates continue to increase with model size. The Qwen 3.5 27B dense reaches 74.7\% and the 35B-A3B MoE reaches 77.3\%.

The best results come from the Qwen 3.6 family, led by the 27B dense at 85.3\% and followed by the 35B-A3B MoE at 81.3\%. Both Qwen 3.6 configurations improve over their Qwen 3.5 counterparts at the same nominal size. Notably, the dense 27B models are substantially slower than the MoE models: Qwen 3.6 27B takes a median of 31.3\,s per attempt, whereas Qwen 3.6 35B-A3B takes 14.7\,s. The MoE architecture therefore provides a strong latency advantage, although the dense Qwen 3.6 27B attains the highest pass rate.

Because task outcomes vary across attempts, Table~\ref{tab:main-results} includes descriptive Wilson 95\% intervals for each pooled pass rate. The point estimates broadly increase with scale, but intervals overlap among the top-performing models (e.g., Qwen 3.6 27B [75.6, 91.6] and Qwen 3.6 35B-A3B [71.1, 88.5]); small differences at the high end should therefore be interpreted cautiously.

\begin{figure}[htbp]
\centering
\includegraphics[width=0.8\linewidth]{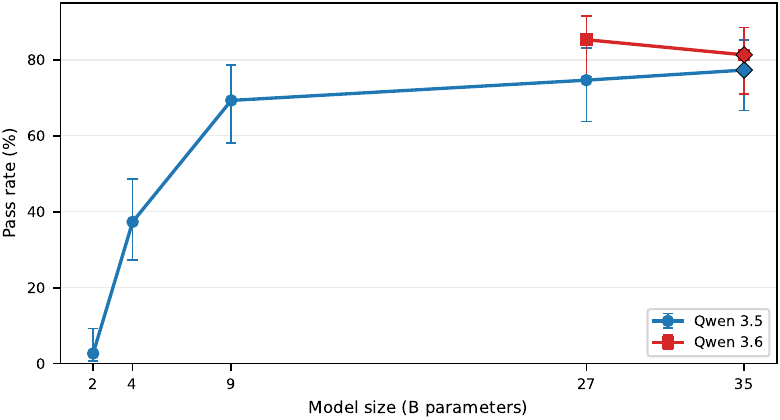}
\caption{Agent-mode pass rate by model size for quantized models. Error bars show Wilson 95\% confidence intervals; lines separate Qwen 3.5 and Qwen 3.6, and diamond markers indicate MoE configurations.}
\label{fig:size-pass-rate}
\end{figure}

These results are also illustrated in Figure~\ref{fig:size-pass-rate}, which plots pass rate against model size, separating the 3.5 and 3.6 families. For Qwen 3.5, the largest improvement occurs between 4B and 9B, after which gains flatten, and the Qwen 3.6 point estimates sit above their Qwen 3.5 counterparts at 27B and 35B-A3B in the evaluated conditions.

We next analyze the observed ``agentic gain'': the increase in pass rate when models can inspect files, use tools, observe execution feedback, and self-verify artifacts instead of receiving one opportunity to submit the result. Table~\ref{tab:agentic-gain} shows that the workspace loop is most valuable for models that can follow the protocol but remain weak in one-shot artifact generation.

The largest gain is observed for Qwen 3.5 9B Q4 (+52.0 percentage points), followed by Qwen 3.5 9B BF16 (+50.7) and Qwen 3.5 4B BF16 (+45.3). The 2B models gain little because they frequently fail to maintain the tool protocol. At the high end, gains remain substantial (+33.3 to +41.3 points), but the relative effect is smaller because the larger models already pass more tasks in \param{chat_only}. Thus, the largest observed gains occur in middle-scale models that can follow the tool-calling protocol but remain weak at one-shot artifact generation.

\begin{table}[htbp]
\centering
\caption{Agentic gain by model.}
\label{tab:agentic-gain}
\scriptsize
\setlength{\tabcolsep}{3pt}
\begin{tabular*}{\linewidth}{@{\extracolsep{\fill}}lccc@{}}
\toprule
Model ID & Chat-only (\%) & Agent-tools (\%) & Agentic gain (pp) \\
\midrule
qwen35\_2b\_q4 & 0.0\% [0.0, 4.9] & 2.7\% [0.7, 9.2] & +2.7 \\
qwen35\_4b\_q4 & 10.7\% [5.5, 19.7] & 37.3\% [27.3, 48.6] & +26.7 \\
qwen35\_9b\_q4 & 17.3\% [10.4, 27.4] & 69.3\% [58.2, 78.6] & +52.0 \\
qwen35\_27b\_q4 & 38.7\% [28.5, 50.0] & 74.7\% [63.8, 83.1] & +36.0 \\
qwen35\_35b\_a3b\_q4 & 40.0\% [29.7, 51.3] & 77.3\% [66.7, 85.3] & +37.3 \\
qwen36\_27b\_q4 & 44.0\% [33.3, 55.3] & 85.3\% [75.6, 91.6] & +41.3 \\
qwen36\_35b\_a3b\_q4 & 48.0\% [37.1, 59.1] & 81.3\% [71.1, 88.5] & +33.3 \\
\midrule
qwen35\_2b\_bf16 & 0.0\% [0.0, 4.9] & 1.3\% [0.2, 7.2] & +1.3 \\
qwen35\_4b\_bf16 & 12.0\% [6.4, 21.3] & 57.3\% [46.1, 67.9] & +45.3 \\
qwen35\_9b\_bf16 & 16.0\% [9.4, 25.9] & 66.7\% [55.4, 76.3] & +50.7 \\
\bottomrule
\end{tabular*}
\end{table}

The mode-specific descriptive intervals do not overlap for any non-2B model in this comparison. We do not treat those endpoints as confidence intervals for the difference between modes. The observed gains in Table~\ref{tab:agentic-gain} characterize the complete workspace condition, not the isolated causal effect of tool access.

The third comparison examines precision within the study's local deployment constraint. All evaluated configurations run on one 24~GB consumer GPU, a setting that supports organizational control over data, model versions, inference settings, and external-service dependencies. Larger models require quantization to fit this memory budget, so the primary scale comparison uses UD-Q4\_K\_XL throughout.

The smaller Qwen 3.5 models also fit in BF16 on the same hardware, permitting an exploratory precision comparison at 2B, 4B, and 9B. Table~\ref{tab:precision-comparison} shows a size-dependent pattern. At 4B, BF16 has a 20.0-point higher agent-mode pass rate than Q4, suggesting that precision may matter near this model's capability threshold. At 9B, Q4 reaches 69.3\% and BF16 reaches 66.7\%, while neither 2B format performs reliably. Because the comparison covers only three sizes and five repetitions per task, it should be treated as evidence of precision sensitivity in these configurations rather than a general conclusion about Q4 quantization.

\begin{table}[htbp]
\centering
\caption{Agent-mode precision comparison for Qwen 3.5 models available in both Q4 and BF16.}
\label{tab:precision-comparison}
\begin{tabular*}{\linewidth}{@{\extracolsep{\fill}}lccc@{}}
\toprule
Size & Q4 pass rate (\%) & BF16 pass rate (\%) & BF16--Q4 (pp) \\
\midrule
2B & 2.7\% [0.7, 9.2] & 1.3\% [0.2, 7.2] & -1.3 \\
4B & 37.3\% [27.3, 48.6] & 57.3\% [46.1, 67.9] & +20.0 \\
9B & 69.3\% [58.2, 78.6] & 66.7\% [55.4, 76.3] & -2.7 \\
\bottomrule
\end{tabular*}
\end{table}

\subsection{Per-task analysis}

The results presented in the previous subsection were aggregated across all tasks. For the sake of completeness, we also present a fine-grained task-level view of the results.

\begin{figure*}[htbp]
\centering
\includegraphics[width=\textwidth]{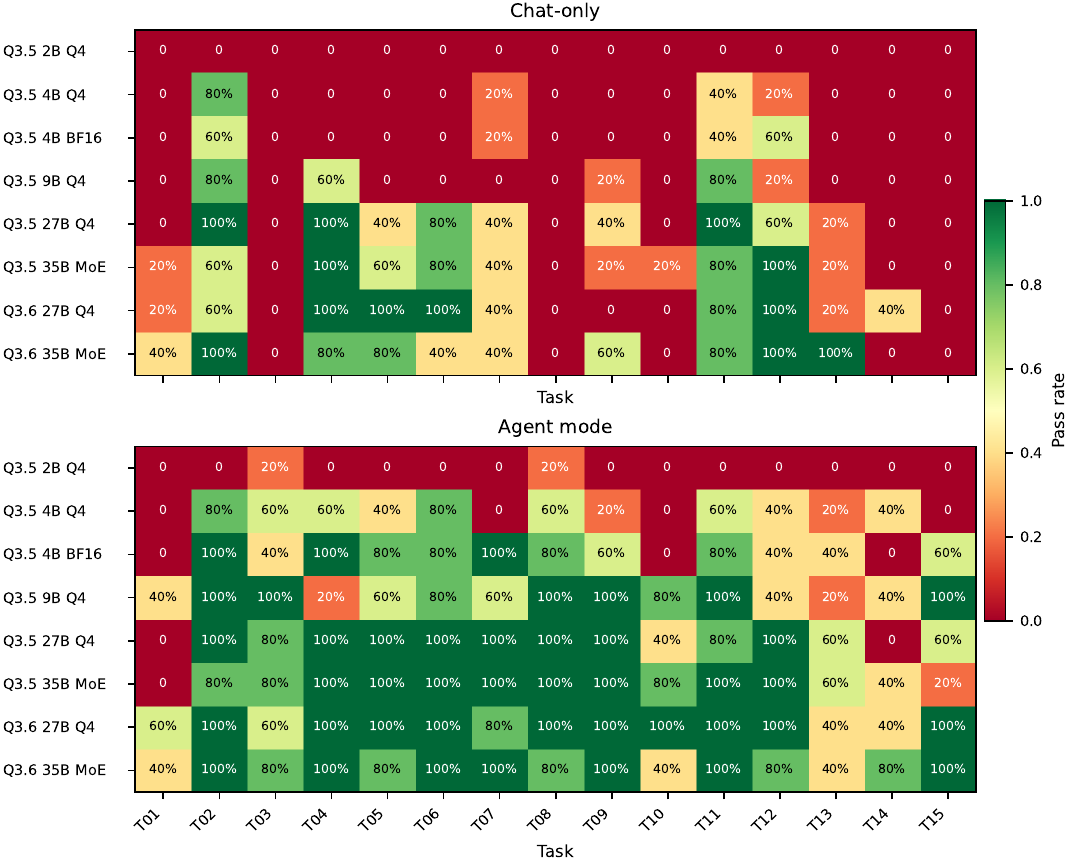}
\caption{Task-level pass rates by mode.}
\label{fig:task-heatmaps}
\end{figure*}

Figure~\ref{fig:task-heatmaps} presents task-level pass rates for chat-only and agent-tools modes. Columns represent tasks, rows represent models, and each cell summarizes five repetitions. The heatmaps include every quantized model and the 4B BF16 comparison; the 2B and 9B BF16 rows are omitted because their aggregate precision differences are small (Table~\ref{tab:precision-comparison}).

The heatmaps expose substantial task-level heterogeneity hidden by aggregate pass rates. No chat-only attempt passes T03, T08, or T15, and T01, T09, T10, and T14 are also difficult in that mode. Agent-mode rates are generally higher, a pattern consistent with the proposed mechanism: visible workspace evidence and revision opportunities allow models to inspect intermediate artifacts and repair some initial mistakes. Because the agent condition changes several interaction features together, the pattern does not establish the effect of any single tool or feedback component.

The 4B BF16 row shows that the aggregate precision difference in Table~\ref{tab:precision-comparison} is heterogeneous across tasks. BF16 performs worse on some cells, including T03, while the Q4 model passes no T07 attempt and the BF16 model passes all five. With only five repetitions per cell, these contrasts identify tasks for follow-up rather than stable task-specific effects.

Finally, the heatmaps also make visible an important reliability issue: each model/task cell is based on five repetitions with the same prompt, inputs, checker, and execution environment. Therefore, an 80\% pass rate means that the model solved four out of five nominally identical attempts, while a 40\% pass rate means that the same model and task alternated between success and failure across repetitions. These intermediate rates show that agents are stochastic systems whose generated plans, tool calls, and repair trajectories can diverge even when the benchmark conditions are fixed. For engineering assessment, a single successful run is therefore weaker evidence than repeated success; repeated trials and task-specific validation are important when the cost of failure is high.

\textbf{Verifiability and agentic gain.} The task-level results also show that agentic gain depends strongly on whether the model can verify its own intermediate work. T08 is the clearest positive example. The task requires converting a Wikidata-style JSON snapshot into a normalized point-of-interest CSV, including coordinate parsing, identifier deduplication, fallback labels, and a fixed output schema. No model passes T08 in \param{chat_only} mode (0/50 attempts), whereas agent mode passes 37/50 attempts. Even Qwen 3.5 2B Q4 passes one repetition in agent mode, Qwen 3.5 4B Q4 passes 3/5, and the Qwen 3.5 9B Q4 and 27B Q4 models pass 5/5. The improvement is not because T08 is conceptually easy in the agent-mode setting; it is because the required transformation is locally inspectable. An agent can open the JSON fixture, write a converter, execute it, inspect the generated CSV, and repair visible problems before submission.

The traces that we inspected support this interpretation. Successful T08 agent runs tend to converge on the same workflow: inspect the actual JSON structure, preserve one output row per unique point-of-interest identifier, handle missing labels consistently, parse coordinates in the expected order, and check the generated file before submitting. Weaker models are less reliable because some repetitions still miss one of these invariants, but successful repetitions show that the feedback loop can compensate when the task exposes concrete, file-level evidence of correctness. In contrast, chat-only attempts cannot observe the fixture or execute the conversion, so they often assume a different JSON layout or produce outputs with invalid row counts or malformed fields. T08 therefore illustrates the setting in which tools provide the strongest benefit: bounded transformations with explicit artifacts and a cheap self-verification path.

T01 illustrates the other side of the same mechanism. The task asks the model to select 10--15 Eurostat indicators relevant to mobility and livability, which requires semantic judgment, theme coverage, and fixture-backed identifier selection. Agent mode helps only modestly on T01 (9/50 attempts versus 4/50 in \param{chat_only}), and among Qwen 3.5 Q4 models the 4B and 27B variants still pass 0/5 agent repetitions while the 9B model passes 2/5.

The harder part is that T01 lacks an equally direct local oracle for semantic adequacy. Reading the indicator file can confirm that a candidate identifier exists, but it does not by itself tell the model whether the selected set satisfies the intended mix of mobility and livability themes, analytical direction, and coverage constraints. The observed failures reflect this gap: some runs omit required themes, some produce malformed CSV structure, and others choose plausible but invalid indicators for the specified selection. Thus, the contrast between T08 and T01 clarifies the main mechanism behind heterogeneous agentic gain. The workspace condition helps most when the model can turn uncertainty into observable checks and repair artifacts from local evidence; it helps less when success depends on semantic grounding that cannot be fully verified through execution or file inspection.

\subsection{Performance analysis}

\begin{figure*}[htbp]
\centering
\includegraphics[width=0.82\textwidth]{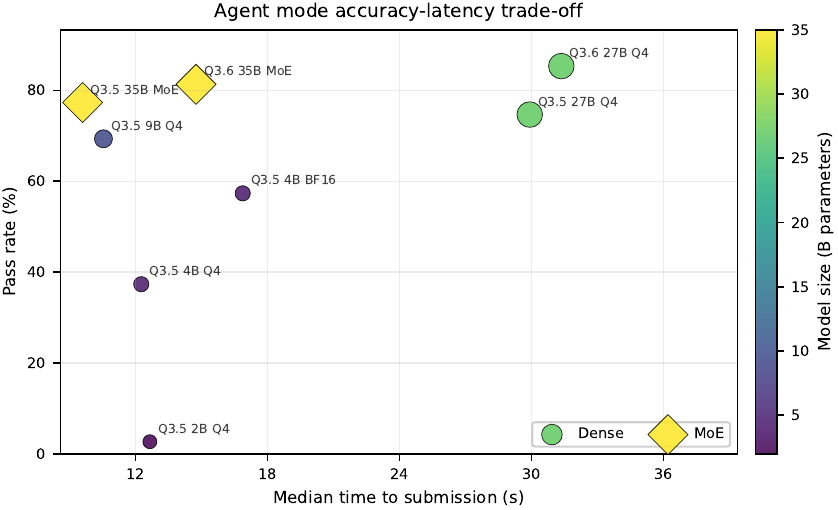}
\caption{Agent-mode artifact-success--latency trade-off for quantized models. The x-axis reports median time to submission and the y-axis reports pass rate; marker color and area encode model size. Dense models are shown with circles and MoE models with diamonds.}
\label{fig:efficiency-scatter}
\end{figure*}

Figure~\ref{fig:efficiency-scatter} compares artifact success with median time to submission for quantized agent-mode runs and the 4B BF16 configuration. Model size is encoded by marker color and area, showing the trade-off between artifact success, scale, and wall-clock latency on the same local hardware.

Both MoE configurations occupy a favorable region relative to the dense 27B models, combining high pass rates with lower median submission times. This pattern is consistent with an efficiency benefit from activated-parameter sparsity, but architecture is not isolated from model configuration in this sweep. The dense Qwen 3.6 27B has the highest pass rate and a higher median latency. The dense 9B Q4 model has a median submission time comparable to the MoE configurations and a smaller memory footprint, making it a candidate for task-specific assessment on constrained hardware despite its lower pass rate.

\begin{figure}[htbp]
\centering
\includegraphics[width=\linewidth]{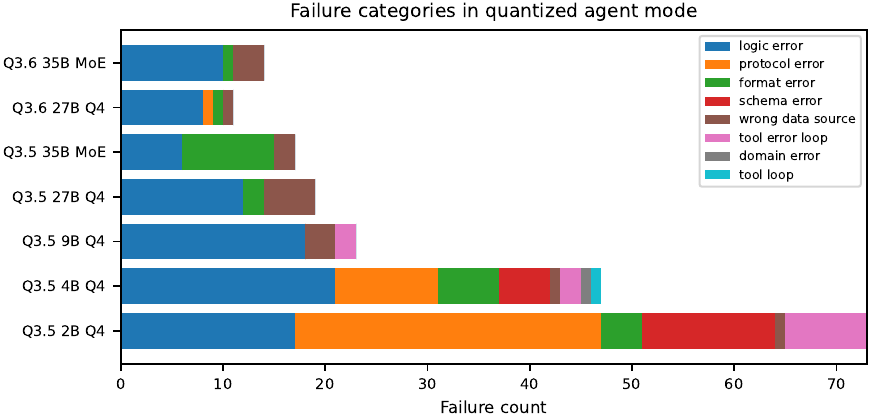}
\caption{Agent-mode failure categories for quantized models. Counts exclude successful attempts and report the raw failure labels.}
\label{fig:agent-failures}
\end{figure}

Figure~\ref{fig:agent-failures} presents failure categories by model. The smallest quantized models are dominated by protocol, schema, and loop failures in agent mode, indicating difficulty operating the tool interface. For larger models, protocol failures mostly disappear and logic errors become the main residual category. Qwen 3.5 35B-A3B has more format errors, including malformed CSV files and reports with missing sections, than its Qwen 3.6 counterpart. The 9B model also exhibits several tool-error loops despite a pass rate closer to the larger models, indicating residual recovery failures at this size.

\section{Discussion}
\label{sec:discussion}

The controlled evaluation shows that local open-weight agents can produce valid artifacts for a substantial subset of the benchmark, but performance is neither uniform nor sufficient for unvalidated automation. Success varies with model scale, the workspace condition, task verifiability, precision, and runtime cost. The following subsections interpret these observations through the four research questions and their implications for LLM-assisted software and data-engineering practice.

\subsection{RQ1: Effect of the workspace interface}

The clearest answer to RQ1 is that the local workspace loop is the most important intervention in the benchmark, provided the model is capable enough to use the interface reliably. Among non-2B models, agent mode improves pass rate by 26.7 to 52.0 percentage points over chat-only mode, making it the largest observed performance shift across the evaluated conditions. The largest absolute gains occur in the middle of the capability range: Qwen 3.5 9B Q4 improves by 52.0 points, Qwen 3.5 9B BF16 by 50.7 points, and Qwen 3.5 4B BF16 by 45.3 points. For larger models, the percentage-point gain is smaller mainly because chat-only performance is higher, but it still reflects the value of the complete workspace loop. At the low end, the 2B models gain little because they do not have enough capability to follow the tool-calling protocol reliably.

The task-level results showed that agentic gain is much larger when the task is self-verifiable, meaning that the model can use local evidence to detect and repair its own mistakes before submission. T08 is the clearest positive case: no model solves it in chat-only mode, whereas agent mode solves 37/50 attempts. The task exposes concrete artifact-level checks, so the workspace loop can turn an initially uncertain transformation into an inspectable and repairable output. T01 shows the opposite boundary. Although the model can inspect the indicator-description fixture, there is no equally direct local oracle for whether a selected indicator set satisfies the intended semantic coverage and analytical-direction constraints. Agent mode helps somewhat overall on T01 (9/50 attempts versus 4/50 in chat-only), but several larger models still fail because the hard part is fixture-grounded domain judgment rather than executable transformation. Thus, the workspace interface is most valuable when tasks expose checkable intermediate states; it is less decisive when the core difficulty is open-ended semantic selection.

\subsection{RQ2: Capability of locally deployable open-weight models}

Artifact success is not restricted to the largest evaluated models. The 2B Q4 model passes only 2/75 agent-mode attempts, and the 4B Q4 model reaches 37.3\%, which is insufficient for dependable use on this suite. The 9B Q4 model reaches 69.3\% and provides a compact comparison point, although its remaining failure rate still requires deterministic validation. Its approximately 6.5~GB checkpoint places validation-gated data-engineering assistance within the memory range of compact local hardware rather than only workstation-class 27B or 35B deployments. Larger local models still improve the frontier: Qwen 3.5 27B Q4 reaches 74.7\%, Qwen 3.5 35B-A3B reaches 77.3\%, Qwen 3.6 35B-A3B reaches 81.3\%, and Qwen 3.6 27B dense reaches the best pass rate at 85.3\%.

The per-task heatmaps show that this capability is not uniform across the mobility workflow. Among quantized 9B+ agent runs, T09 is solved in every attempt, and T02, T06, T08, and T11 all reach at least 96\%. These tasks are procedural, fixture-grounded, and easy to inspect through generated artifacts. In contrast, T01, T13, and T14 remain difficult. The harder tasks require semantic selection, exact clustering behavior, or complete validation logic, which are less easily repaired by generic execution feedback.

The failure taxonomy is consistent with this view: stronger local agents fail less often because of interface-level breakdowns and more often because of task-specific reasoning, validation, or fixture-grounding mistakes.

The format-error pattern adds a useful nuance to this interpretation. Once a model can operate the workspace loop, the remaining failure mode is not always incorrect reasoning in the narrow sense; it can also be insufficient final validation of otherwise plausible artifacts. The Qwen 3.5 35B-A3B failures illustrate this point: several attempts reached submission with files or reports in the expected location, but the artifacts still violated parser-level or section-level requirements. In practice, this suggests that stronger local agents may benefit less from more raw tool access and more from explicit self-checking routines before submission, such as re-reading generated CSVs with the same library used downstream or verifying required report headings.

\subsection{RQ3: Artifact success, runtime, and tool-use trade-offs}

The efficiency results show that deployment decisions cannot be based on pass rate alone. Higher capability generally increases pass rate, but the most accurate model is not always the most attractive deployment choice. Qwen 3.6 27B dense has the highest pass rate at 85.3\%, but it also has the highest median submission time at 31.3s. The MoE models occupy a more latency-efficient region of the artifact-success--latency plot: Qwen 3.5 35B-A3B reaches 77.3\% at 9.6s, and Qwen 3.6 35B-A3B reaches 81.3\% at 14.7s. The MoE configurations therefore show a practical wall-clock advantage in this local setting, although the design does not isolate architecture as its sole cause and the dense Qwen 3.6 27B remains the pass-rate leader.

As previously mentioned, the 9B Q4 model is especially relevant for this deployment trade-off. Its pass rate is lower than the 27B and 35B-class models, but it reaches 69.3\% while occupying roughly 6.5~GB and matching the MoE models in median submission time. This makes it a candidate for task-specific evaluation on compact machines or privacy-sensitive workflows where memory footprint and responsiveness matter alongside artifact correctness. In contrast, the larger models have higher pass rates but require a workstation-class GPU budget. The engineering choice is therefore a position on the artifact-success/latency/memory frontier rather than a single universally best model.

\subsection{RQ4: Effect of quantization and precision}

The precision comparison shows that quantization is not uniformly harmful or harmless; its effect depends on where the base model sits relative to the capability threshold. BF16 does not rescue the 2B model, and it does not improve the 9B model, where Q4 is already slightly ahead within the sampling variability of five repetitions. At 4B, however, BF16 increases pass rate from 37.3\% to 57.3\%. The most plausible interpretation is that the 4B model lies close to the minimum capability boundary: losing precision pushes many attempts below the threshold for reliable parsing, tool use, and artifact construction, while BF16 gives enough headroom for more tasks to complete.

Once the base model is sufficiently capable, Q4 quantization appears viable for this benchmark. The 9B Q4 model is more accurate, smaller and faster than the 4B BF16 model, and the best overall models in the sweep are quantized larger models served within the single-24~GB-GPU constraint. The 4B BF16 point still has practical value because it shows that precision can matter near the threshold, but it does not dominate the larger quantized alternatives. Within this benchmark, aggressive quantization is therefore not merely a mechanism for fitting larger models into memory; some quantized configurations retain substantial task performance. Broader claims require replication across model families, tasks, and quantization methods.

The Qwen 3.5--3.6 comparison adds one final qualification. At the two matched sizes, Qwen 3.6 has higher point estimates: 85.3\% versus 74.7\% for 27B dense and 81.3\% versus 77.3\% for 35B-A3B. In this sweep, 9B Q4 is the first configuration to pass a majority of attempts, 35B-A3B offers the lowest latency among the higher-performing configurations, and Qwen 3.6 27B dense has the highest pass rate. Selection should nevertheless be based on representative tasks and jointly consider artifact correctness, latency, memory footprint, and protocol robustness.

\subsection{Threats to validity}
\label{sec:threats-to-validity}

The controlled design supports consistent comparisons across the evaluated configurations, while the following considerations define the scope in which the findings should be interpreted.

\textbf{Construct validity.} Artifact success captures the dimensions targeted by the benchmark, including execution, schema conformance, numerical consistency, report structure, and selected domain rules. This focus matches the study objective of assessing whether agents can deliver artifacts that satisfy explicit engineering requirements. It does not directly assess broader qualities such as maintainability, security, or subjective prose quality. Runtime and tool-call counts describe the tested harness and hardware.

\textbf{Internal validity.} The two modes represent complete interaction configurations. Because the workspace condition combines inspection, execution feedback, tool use, and revision, the observed gain is attributed to the configuration rather than to one component. Prompts, fixtures, checkers, generation settings, and workspace constraints were held fixed to support this comparison. Five repetitions per cell capture recurring variability, although rare generation trajectories may not appear.

\textbf{External validity.} The controlled Qwen-family sweep, common serving stack, and single GPU support consistent within-study comparison. The fifteen tasks span five workflow stages and several artifact types, providing broader coverage than a single code-generation task. Results may nevertheless vary with other model families, agent designs, hardware, domains, and production-scale pipelines, which provides a natural direction for replication.

\textbf{Conclusion validity.} Aggregate rates combine tasks of different difficulty, and Wilson intervals summarize the observed attempts descriptively rather than supporting population-level inference. The matched Q4--BF16 comparison covers three model sizes and is therefore exploratory. We consequently emphasize effect sizes and task-level patterns. Repeated attempts and released traces complement the aggregate analysis, while the replication package supports broader follow-up studies.

\section{Conclusion}
\label{sec:conclusion}

This paper introduced a reproducible, artifact-level benchmark for evaluating local LLM agents as data-engineering assistants, using mobility workflows as a domain-grounded empirical testbed. Across 1,500 attempts, the evaluated configurations exhibited substantial differences in artifact correctness, interaction reliability, runtime, and failure behavior. The findings show that local agents can support parts of software-intensive data-engineering work, but they do not justify unvalidated autonomous execution.

The largest observed performance shift is associated with the complete workspace condition. Among non-2B models, pass rates are 26.7--52.0 percentage points higher than under one-shot generation. The comparison does not isolate tool access from iterative context or revision, but it demonstrates the engineering value of a configuration in which models can inspect files, execute code, observe errors, and repair artifacts. Gains are greatest when intermediate states are locally verifiable through parsers, executable scripts, numerical checks, or inspectable files.

The model sweep also identifies a resource--correctness trade-off. Qwen 3.6 27B dense attains the highest pass rate at 85.3\%, while Qwen 3.5 9B Q4 reaches 69.3\% with an approximately 6.5~GB checkpoint. The smaller configuration may be suitable for validation-gated assistance on constrained hardware, whereas the larger models reduce failures at higher memory or latency cost. None of the measured pass rates removes the need for task-specific checkers.

MoE configurations occupy a favorable artifact-success--latency region, and the exploratory precision comparison suggests that BF16 matters more at 4B than at 9B in this study. These observations are configuration-specific rather than general claims about architecture or quantization. More broadly, selecting an engineering agent requires evaluating artifact correctness together with latency, memory footprint, protocol reliability, and the availability of deterministic validation.

The benchmark and replication package provide a basis for extending this empirical evaluation to additional model families, agent scaffolds, hardware, and task suites. For software and data-engineering teams, the central implication is not that a particular local model is universally ready for deployment, but that complete agent configurations can be evaluated reproducibly through the artifacts they produce. Workflows with explicit specifications and executable checks offer the strongest setting for validation-gated LLM assistance.

\section*{CRediT authorship contribution statement}
\textbf{Jorge García-Carrasco}: Conceptualization, Methodology, Software, Validation, Investigation, Data curation, Writing -- original draft, Writing -- review \& editing, Visualization.
\textbf{Javier Sanchis}: Conceptualization, Writing -- review \& editing.
\textbf{Alejandro Reina-Reina}: Writing -- review \& editing.
\textbf{Alejandro Maté}: Supervision, Project administration, Conceptualization, Writing -- review \& editing.
\textbf{Juan Trujillo}: Supervision, Project administration, Conceptualization, Writing -- review \& editing.

\section*{Funding}
This work was supported by the European Union NextGenerationEU through the Spanish Ministry for Digital Transformation and the Civil Service under the SOPHIA project [TSI-100130-2024-10] and the ENIA Chair of Artificial Intelligence at the University of Alicante [TSI-100927-2023-6]; the Spanish Ministry of Science and Innovation under the KOSMOS-UA project [PID2024-155363OB-C43]; Generalitat Valenciana under the BALIDA-AA [CIPROM/2024/13] and IAEAV [INREIA/2024/176] projects; and MCIN/\hspace{0pt}AEI [RED2022-134656-T]. The funding bodies had no role in the study design, data collection, analysis, interpretation, manuscript preparation, or decision to submit.

\section*{Data availability}
The versioned replication package containing the benchmark code, fixed data fixtures, scored attempts, execution traces, provenance records, aggregate results, and analysis scripts is archived on Zenodo \citep{garciacarrasco2026replication}.

\section*{Declaration of competing interests}
The authors declare that they have no known competing financial interests or personal relationships that could have appeared to influence the work reported in this paper.

\bibliographystyle{unsrtnat}
\bibliography{references}

@misc{eurostatApi2026,
  author = {{Eurostat}},
  title = {Dissemination {API}: Statistics Endpoint},
  year = {2026},
  url = {https://ec.europa.eu/eurostat/api/dissemination/statistics/1.0/data/},
  note = {Accessed 27 April 2026}
}

@misc{dataEuropaApi2026,
  author = {{data.europa.eu}},
  title = {Search {API}},
  year = {2026},
  url = {https://data.europa.eu/api/hub/search/search},
  note = {Accessed 27 April 2026}
}

@misc{mobilityDataCatalog2026,
  author = {{MobilityData}},
  title = {Mobility Database Catalogs},
  year = {2026},
  url = {https://files.mobilitydatabase.org/feeds_v2.csv},
  note = {Accessed 27 April 2026}
}

@misc{wikidataSparql2026,
  author = {{Wikidata}},
  title = {{Wikidata} Query Service {SPARQL} Endpoint},
  year = {2026},
  url = {https://query.wikidata.org/sparql},
  note = {Accessed 27 April 2026}
}

@inproceedings{yao2023react,
  title = {{ReAct}: Synergizing Reasoning and Acting in Language Models},
  author = {Yao, Shunyu and Zhao, Jeffrey and Yu, Dian and Du, Nan and Shafran, Izhak and Narasimhan, Karthik R. and Cao, Yuan},
  booktitle = {Proceedings of the 11th International Conference on Learning Representations},
  year = {2023},
  url = {https://openreview.net/forum?id=WE_vluYUL-X}
}

@inproceedings{schick2023toolformer,
  title = {{Toolformer}: Language Models Can Teach Themselves to Use Tools},
  author = {Schick, Timo and Dwivedi-Yu, Jane and Dess{\`i}, Roberto and Raileanu, Roberta and Lomeli, Maria and Hambro, Eric and Zettlemoyer, Luke and Cancedda, Nicola and Scialom, Thomas},
  booktitle = {Advances in Neural Information Processing Systems},
  volume = {36},
  year = {2023},
  url = {https://proceedings.neurips.cc/paper_files/paper/2023/hash/d842425e4bf79ba039352da0f658a906-Abstract-Conference.html}
}

@article{chen2021codex,
  title = {Evaluating Large Language Models Trained on Code},
  author = {Chen, Mark and Tworek, Jerry and Jun, Heewoo and Yuan, Qiming and Pinto, Henrique Ponde de Oliveira and Kaplan, Jared and Edwards, Harri and Burda, Yuri and Joseph, Nicholas and Brockman, Greg and Ray, Alex and Puri, Raul and Krueger, Gretchen and Petrov, Michael and Khlaaf, Heidy and Sastry, Girish and Mishkin, Pamela and Chan, Brooke and Gray, Scott and Ryder, Nick and Pavlov, Mikhail and Power, Alethea and Kaiser, Lukasz and Bavarian, Mohammad and Winter, Clemens and Tillet, Philippe and Such, Felipe Petroski and Cummings, Dave and Plappert, Matthias and Chantzis, Fotios and Barnes, Elizabeth and Herbert-Voss, Ariel and Guss, William Hebgen and Nichol, Alex and Paino, Alex and Tezak, Nikolas and Tang, Jie and Babuschkin, Igor and Balaji, Suchir and Jain, Shantanu and Saunders, William and Hesse, Christopher and Carr, Andrew N. and Leike, Jan and Achiam, Josh and Misra, Vedant and Morikawa, Evan and Radford, Alec and Knight, Matthew and Brundage, Miles and Murati, Mira and Mayer, Katie and Welinder, Peter and McGrew, Bob and Amodei, Dario and McCandlish, Sam and Sutskever, Ilya and Zaremba, Wojciech},
  journal = {arXiv preprint arXiv:2107.03374},
  year = {2021},
  url = {https://arxiv.org/abs/2107.03374}
}

@article{austin2021programsynthesis,
  title = {Program Synthesis with Large Language Models},
  author = {Austin, Jacob and Odena, Augustus and Nye, Maxwell and Bosma, Maarten and Michalewski, Henryk and Dohan, David and Jiang, Ellen and Cai, Carrie and Terry, Michael and Le, Quoc and Sutton, Charles},
  journal = {arXiv preprint arXiv:2108.07732},
  year = {2021},
  url = {https://arxiv.org/abs/2108.07732}
}

@inproceedings{lu2021codexglue,
  title = {{CodeXGLUE}: A Machine Learning Benchmark Dataset for Code Understanding and Generation},
  author = {Lu, Shuai and Guo, Daya and Ren, Shuo and Huang, Junjie and Svyatkovskiy, Alexey and Blanco, Ambrosio and Clement, Colin and Drain, Dawn and Jiang, Daxin and Tang, Duyu and Li, Ge and Zhou, Lidong and Shou, Linjun and Zhou, Long and Tufano, Michele and Gong, Ming and Zhou, Ming and Duan, Nan and Sundaresan, Neel and Deng, Shao Kun and Fu, Shengyu and Liu, Shujie},
  booktitle = {Proceedings of the Neural Information Processing Systems Track on Datasets and Benchmarks},
  year = {2021},
  url = {https://datasets-benchmarks-proceedings.neurips.cc/paper/2021/hash/c16a5320fa475530d9583c34fd356ef5-Abstract-round1.html}
}

@inproceedings{lai2023ds1000,
  title = {{DS-1000}: A Natural and Reliable Benchmark for Data Science Code Generation},
  author = {Lai, Yuhang and Li, Chengxi and Wang, Yiming and Zhang, Tianyi and Zhong, Ruiqi and Zettlemoyer, Luke and Yih, Wen-tau and Fried, Daniel and Wang, Sida and Yu, Tao},
  booktitle = {Proceedings of the 40th International Conference on Machine Learning},
  series = {Proceedings of Machine Learning Research},
  volume = {202},
  year = {2023},
  publisher = {PMLR},
  url = {https://proceedings.mlr.press/v202/lai23b.html}
}

@inproceedings{jimenez2024swebench,
  title = {{SWE-bench}: Can Language Models Resolve Real-world {GitHub} Issues?},
  author = {Jimenez, Carlos E. and Yang, John and Wettig, Alexander and Yao, Shunyu and Pei, Kexin and Press, Ofir and Narasimhan, Karthik},
  booktitle = {Proceedings of the 12th International Conference on Learning Representations},
  year = {2024},
  url = {https://openreview.net/forum?id=VTF8yNQM66}
}

@inproceedings{li2023apibank,
  title = {{API}-Bank: A Comprehensive Benchmark for Tool-Augmented {LLM}s},
  author = {Li, Minghao and Zhao, Yingxiu and Yu, Bowen and Song, Feifan and Li, Hangyu and Yu, Haiyang and Li, Zhoujun and Huang, Fei and Li, Yongbin},
  booktitle = {Proceedings of the 2023 Conference on Empirical Methods in Natural Language Processing},
  pages = {3102--3116},
  year = {2023},
  address = {Singapore},
  publisher = {Association for Computational Linguistics},
  doi = {10.18653/v1/2023.emnlp-main.187},
  url = {https://aclanthology.org/2023.emnlp-main.187/}
}

@inproceedings{liu2024agentbench,
  title = {{AgentBench}: Evaluating {LLM}s as Agents},
  author = {Liu, Xiao and Yu, Hao and Zhang, Hanchen and Xu, Yifan and Lei, Xuanyu and Lai, Hanyu and Gu, Yu and Ding, Hangliang and Men, Kaiwen and Yang, Kejuan and Zhang, Shudan and Deng, Xiang and Zeng, Aohan and Du, Zhengxiao and Zhang, Chenhui and Shen, Sheng and Zhang, Tianjun and Su, Yu and Sun, Huan and Huang, Minlie and Dong, Yuxiao and Tang, Jie},
  booktitle = {Proceedings of the 12th International Conference on Learning Representations},
  year = {2024},
  url = {https://openreview.net/forum?id=zAdUB0aCTQ}
}

@inproceedings{guo2024stabletoolbench,
  title = {{StableToolBench}: Towards Stable Large-Scale Benchmarking on Tool Learning of Large Language Models},
  author = {Guo, Zhicheng and Cheng, Sijie and Wang, Hao and Liang, Shihao and Qin, Yujia and Li, Peng and Liu, Zhiyuan and Sun, Maosong and Liu, Yang},
  booktitle = {Findings of the Association for Computational Linguistics: ACL 2024},
  pages = {11143--11156},
  year = {2024},
  address = {Bangkok, Thailand},
  publisher = {Association for Computational Linguistics},
  doi = {10.18653/v1/2024.findings-acl.664},
  url = {https://aclanthology.org/2024.findings-acl.664/}
}

@inproceedings{patil2025bfcl,
  title = {The {Berkeley Function Calling Leaderboard} ({BFCL}): From Tool Use to Agentic Evaluation of Large Language Models},
  author = {Patil, Shishir G. and Mao, Huanzhi and Yan, Fanjia and Ji, Charlie Cheng-Jie and Suresh, Vishnu and Stoica, Ion and Gonzalez, Joseph E.},
  booktitle = {Proceedings of the 42nd International Conference on Machine Learning},
  series = {Proceedings of Machine Learning Research},
  volume = {267},
  pages = {48371--48392},
  year = {2025},
  publisher = {PMLR},
  url = {https://proceedings.mlr.press/v267/patil25a.html}
}

@article{yao2024taubench,
  title = {{$\tau$-bench}: A Benchmark for Tool-Agent-User Interaction in Real-World Domains},
  author = {Yao, Shunyu and Shinn, Noah and Razavi, Pedram and Narasimhan, Karthik},
  journal = {arXiv preprint arXiv:2406.12045},
  year = {2024},
  url = {https://arxiv.org/abs/2406.12045}
}

@inproceedings{lu2025toolsandbox,
  title = {{ToolSandbox}: A Stateful, Conversational, Interactive Evaluation Benchmark for {LLM} Tool Use Capabilities},
  author = {Lu, Jiarui and Holleis, Thomas and Zhang, Yizhe and Aumayer, Bernhard and Nan, Feng and Bai, Haoping and Ma, Shuang and Ma, Shen and Li, Mengyu and Yin, Guoli and Wang, Zirui and Pang, Ruoming},
  booktitle = {Findings of the Association for Computational Linguistics: NAACL 2025},
  pages = {1160--1183},
  year = {2025},
  publisher = {Association for Computational Linguistics},
  doi = {10.18653/v1/2025.findings-naacl.65},
  url = {https://aclanthology.org/2025.findings-naacl.65/}
}

@inproceedings{frantar2023gptq,
  title = {{GPTQ}: Accurate Post-Training Quantization for Generative Pre-trained Transformers},
  author = {Frantar, Elias and Ashkboos, Saleh and Hoefler, Torsten and Alistarh, Dan},
  booktitle = {Proceedings of the 11th International Conference on Learning Representations},
  year = {2023},
  url = {https://openreview.net/forum?id=tcbBPnfwxS}
}

@inproceedings{lin2024awq,
  title = {{AWQ}: Activation-Aware Weight Quantization for On-Device {LLM} Compression and Acceleration},
  author = {Lin, Ji and Tang, Jiaming and Tang, Haotian and Yang, Shang and Chen, Wei-Ming and Wang, Wei-Chen and Xiao, Guangxuan and Dang, Xingyu and Gan, Chuang and Han, Song},
  booktitle = {Proceedings of Machine Learning and Systems},
  year = {2024},
  url = {https://mlsys.org/virtual/2024/poster/2653}
}

@inproceedings{gundersen2018reproducibility,
  title = {State of the Art: Reproducibility in Artificial Intelligence},
  author = {Gundersen, Odd Erik and Kjensmo, Sigbj{\o}rn},
  booktitle = {Proceedings of the AAAI Conference on Artificial Intelligence},
  volume = {32},
  number = {1},
  year = {2018},
  doi = {10.1609/aaai.v32i1.11503},
  url = {https://aaai.org/papers/11503-state-of-the-art-reproducibility-in-artificial-intelligence/}
}

@article{pineau2021reproducibility,
  title = {Improving Reproducibility in Machine Learning Research: A Report from the {NeurIPS} 2019 Reproducibility Program},
  author = {Pineau, Joelle and Vincent-Lamarre, Philippe and Sinha, Koustuv and Larivi{\`e}re, Vincent and Beygelzimer, Alina and d'Alch{\'e}-Buc, Florence and Fox, Emily and Larochelle, Hugo},
  journal = {Journal of Machine Learning Research},
  volume = {22},
  number = {164},
  pages = {1--20},
  year = {2021},
  url = {https://www.jmlr.org/papers/v22/20-303.html}
}

@inproceedings{mazumder2023dataperf,
  title = {{DataPerf}: Benchmarks for Data-Centric {AI} Development},
  author = {Mazumder, Mark and Banbury, Colby and Yao, Xiaozhe and Karlas, Bojan and Rojas, William Gaviria and Diamos, Sudnya and Diamos, Greg and He, Lynn and Parrish, Alicia and Kirk, Hannah Rose and Quaye, Jessica and Rastogi, Charvi and Kiela, Douwe and Jurado, David and Kanter, David and Mosquera, Rafael and Cukierski, Will and Ciro, Juan and Aroyo, Lora and Acun, Bilge and Chen, Lingjiao and Raje, Mehul and Bartolo, Max and Eyuboglu, Evan Sabri and Ghorbani, Amirata and Goodman, Emmett and Howard, Addison and Inel, Oana and Kane, Tariq and Kirkpatrick, Christine R. and Sculley, D. and Kuo, Tzu-Sheng and Mueller, Jonas W. and Thrush, Tristan and Vanschoren, Joaquin and Warren, Margaret and Williams, Adina and Yeung, Serena and Ardalani, Newsha and Paritosh, Praveen and Zhang, Ce and Zou, James Y. and Wu, Carole-Jean and Coleman, Cody and Ng, Andrew Y. and Mattson, Peter and Reddi, Vijay Janapa},
  booktitle = {Advances in Neural Information Processing Systems},
  volume = {36},
  year = {2023},
  url = {https://proceedings.neurips.cc/paper_files/paper/2023/hash/112db88215e25b3ae2750e9eefcded94-Abstract-Datasets_and_Benchmarks.html}
}

@misc{llamacpp2026,
  title = {{llama.cpp}: {LLM} Inference in {C/C++}},
  author = {{ggml-org}},
  year = {2026},
  howpublished = {\url{https://github.com/ggml-org/llama.cpp}},
  note = {Accessed 28 April 2026}
}

@misc{gtfsScheduleReference2026,
  title = {{GTFS} Schedule Reference},
  author = {{MobilityData}},
  year = {2026},
  howpublished = {\url{https://gtfs.org/documentation/schedule/reference/}},
  note = {Accessed 28 April 2026}
}

@misc{mobilityDataValidator2026,
  title = {{GTFS} Schedule Validator Rules},
  author = {{MobilityData}},
  year = {2026},
  howpublished = {\url{https://gtfs-validator.mobilitydata.org/rules.html}},
  note = {Accessed 28 April 2026}
}

@misc{qwen3.6-27b,
  title = {{Qwen3.6-27B}: Flagship-Level Coding in a 27B Dense Model},
  author = {{Qwen Team}},
  year = {2026},
  month = {April},
  howpublished = {\url{https://qwen.ai/blog?id=qwen3.6-27b}},
  note = {Accessed 28 April 2026}
}

@misc{unsloth_quant,
  author       = {{Unsloth}},
  title        = {{Unsloth Dynamic 2.0 GGUFs}},
  year         = {2026},
  howpublished = {\url{https://unsloth.ai/docs/basics/unsloth-dynamic-2.0-ggufs}},
  note         = {Accessed 5 May 2026}
}

@inproceedings{shazeer2017outrageously,
  title     = {Outrageously Large Neural Networks: The Sparsely-Gated Mixture-of-Experts Layer},
  author    = {Shazeer, Noam and Mirhoseini, Azalia and Maziarz, Krzysztof and Davis, Andy and Le, Quoc V. and Hinton, Geoffrey E. and Dean, Jeff},
  booktitle = {International Conference on Learning Representations},
  year      = {2017},
  url       = {https://openreview.net/forum?id=B1ckMDqlg}
}

@inproceedings{hollmann2023large,
  title     = {Large Language Models for Automated Data Science: Introducing {CAAFE} for Context-Aware Automated Feature Engineering},
  author    = {Hollmann, Noah and M{\"u}ller, Samuel and Hutter, Frank},
  booktitle = {Advances in Neural Information Processing Systems},
  volume    = {36},
  pages     = {44753--44775},
  year      = {2023}
}

@article{jiang2026survey,
  title     = {A Survey on Large Language Models for Code Generation},
  author    = {Jiang, Juyong and Wang, Fan and Shen, Jiasi and Kim, Sungju and Kim, Sunghun},
  journal   = {{ACM} Transactions on Software Engineering and Methodology},
  volume    = {35},
  number    = {2},
  articleno = {58},
  pages     = {1--72},
  year      = {2026},
  doi       = {10.1145/3747588}
}

@inproceedings{feldt2023towards,
  title     = {Towards Autonomous Testing Agents via Conversational Large Language Models},
  author    = {Feldt, Robert and Kang, Sungmin and Yoon, Juyeon and Yoo, Shin},
  booktitle = {Proceedings of the 38th {IEEE/ACM} International Conference on Automated Software Engineering},
  series    = {ASE 2023},
  pages     = {1688--1693},
  year      = {2023},
  doi       = {10.1109/ASE56229.2023.00148}
}

@article{wang2024survey,
  title     = {A Survey on Large Language Model Based Autonomous Agents},
  author    = {Wang, Lei and Ma, Chen and Feng, Xueyang and Zhang, Zeyu and Yang, Hao and Zhang, Jingsen and Chen, Zhiyuan and Tang, Jiakai and Chen, Xu and Lin, Yankai and Zhao, Wayne Xin and Wei, Zhewei and Wen, Jirong},
  journal   = {Frontiers of Computer Science},
  volume    = {18},
  number    = {6},
  pages     = {186345},
  year      = {2024},
  doi       = {10.1007/s11704-024-40231-1}
}

@article{pornprasit2024codereview,
  author = {Pornprasit, Chanathip and Tantithamthavorn, Chakkrit},
  title = {Fine-Tuning and Prompt Engineering for Large Language Models-Based Code Review Automation},
  journal = {Information and Software Technology},
  volume = {175},
  pages = {107523},
  year = {2024},
  doi = {10.1016/j.infsof.2024.107523},
  url = {https://doi.org/10.1016/j.infsof.2024.107523}
}

@article{nguyen2025codesemantics,
  author = {Nguyen, Thu-Trang and Vu, Thanh Trong and Vo, Hieu Dinh and Nguyen, Son},
  title = {An Empirical Study on Capability of Large Language Models in Understanding Code Semantics},
  journal = {Information and Software Technology},
  volume = {185},
  pages = {107780},
  year = {2025},
  doi = {10.1016/j.infsof.2025.107780},
  url = {https://doi.org/10.1016/j.infsof.2025.107780}
}

@article{liu2026ucdllm,
  author = {Liu, Xianhui and Liu, Yueying and Zhuang, Yiheng and Hou, Wenlong},
  title = {{UCD-LLM}: A Use Case Diagram Requirement Modeling Multi-Agent Framework with Large Language Model},
  journal = {Information and Software Technology},
  volume = {190},
  pages = {107955},
  year = {2026},
  doi = {10.1016/j.infsof.2025.107955},
  url = {https://doi.org/10.1016/j.infsof.2025.107955}
}

@article{kim2026artifactvalidity,
  author = {Kim, Dae-Kyoo},
  title = {Artifact Validity under Varying Agent Configurations in {LLM}-Assisted Software Development: A Comparative Analysis},
  journal = {Information and Software Technology},
  volume = {192},
  pages = {108022},
  year = {2026},
  doi = {10.1016/j.infsof.2026.108022},
  url = {https://doi.org/10.1016/j.infsof.2026.108022}
}

@misc{garciacarrasco2026replication,
  author = {García-Carrasco, Jorge and Sanchis, Javier and Reina-Reina, Alejandro and Maté, Alejandro and Trujillo, Juan},
  title = {Evaluating Local Language Model Agents for Reproducible Data Engineering: An Empirical Software Engineering Study of Mobility Workflows},
  year = {2026},
  publisher = {Zenodo},
  version = {v1.0.0},
  doi = {10.5281/zenodo.21397610},
  url = {https://doi.org/10.5281/zenodo.21397610}
}
\end{document}